# An evolutionary continuum from nucleated dwarf galaxies to star clusters

Kaixiang Wang[1,2,*], Eric W. Peng[3,*], Chengze Liu[4,5,*], J. Christopher Mihos[6], Patrick Côté[7], Laura Ferrarese[7], Matthew A. Taylor[8], John P. Blakeslee[3], Jean-Charles Cuillandre[9], Pierre-Alain Duc[10], Puragra Guhathakurta[11], Stephen Gwyn[7], Youkyung Ko[12], Ariane Lançon[10], Sungsoon Lim[13], Lauren A. MacArthur[14], Thomas Puzia[15], Joel Roediger[7], Laura V. Sales[16], Rubén Sánchez-Janssen[17], Chelsea Spengler[15], Elisa Toloba[18], Hongxin Zhang[19,20], and Mingcheng Zhu[1,2]

[1]Department of Astronomy, Peking University, Beijing 100871, China
[2]Kavli Institute for Astronomy and Astrophysics, Peking University, Beijing 100871, China
[3]NSF's NOIRLab, 950 N. Cherry Ave, Tucson, AZ 85719, USA
[4]Department of Astronomy, School of Physics and Astronomy, and Shanghai Key Laboratory for Particle Physics and Cosmology, Shanghai Jiao Tong University, Shanghai 200240, China
[5]Tsung-Dao Lee Institute, Shanghai Jiao Tong University, Shengrong Road 520, Shanghai, 201210, China
[6]Department of Astronomy, Case Western Reserve University, 10900 Euclid Avenue, Cleveland, OH 44106, USA
[7]Herzberg Astronomy and Astrophysics Research Centre, National Research Council of Canada, 5071 W. Saanich Road, Victoria, BC, V9E 2E7, Canada
[8]Department of Physics and Astronomy, University of Calgary, 2500 University Drive NW, Calgary, AB, T2N 1N4, Canada
[9]AIM Paris Saclay, CNRS/INSU, CEA/Irfu, Université Paris Diderot, Orme des Merisiers, F-91191 Gif-sur-Yvette Cedex, France
[10]Université de Strasbourg, CNRS, Observatoire astronomique de Strasbourg, UMR 7550, Strasbourg, France
[11]UCO/Lick Observatory, Department of Astronomy and Astrophysics, University of California Santa Cruz, 1156 High Street, Santa Cruz, CA 95064, USA
[12]Korea Astronomy and Space Science Institute, 776 Daedeok-daero, Yuseong-Gu, Daejeon 34055, Republic of Korea
[13]Department of Astronomy, Yonsei University, 50 Yonsei-ro, Seodaemun-gu, Seoul 03722, Republic of Korea
[14]Department of Astrophysical Sciences, Princeton University, Princeton, New Jersey 08544, USA
[15]Institute of Astrophysics, Pontificia Universidad Católica de Chile, Av. Vicuña Mackenna 4860, 7820436 Macul, Santiago, Chile
[16]Department of Physics and Astronomy, University of California, Riverside, CA 92521, USA
[17]UK Astronomy Technology Centre, Royal Observatory Edinburgh, Blackford Hill, Edinburgh EH9 3HJ, UK
[18]Department of Physics and Astronomy, University of the Pacific, 3601 Pacific Avenue, Stockton, CA 95211, USA
[19]School of Astronomy and Space Science, University of Science and Technology of China, Hefei 230026, China
[20]CAS Key Laboratory for Research in Galaxies and Cosmology, Department of Astronomy, University of Science and Technology of China, Hefei, Anhui 230026, China

*Corresponding Author, kaixiang.wang@pku.edu.cn
*Corresponding Author, eric.peng@noirlab.edu
*Corresponding Author, czliu@sjtu.edu.cn

**Systematic studies[1-4] have revealed hundreds of ultra-compact dwarf galaxies (UCDs[5]) in the nearby Universe. With half-light radii $r_h$ of approximately 10-100 parsecs and stellar masses $M_* \approx 10^6$–$10^8$ solar masses, UCDs are among the densest known stellar systems[6]. Although similar in appearance to massive globular clusters[7], the detection of extended stellar envelopes[4,8,9], complex star formation histories[10], elevated mass-to-light ratio[11,12], and supermassive black holes[13-16] suggest that some UCDs are remnant nuclear star clusters[17] of tidally-stripped dwarf galaxies[18-19], or even ancient compact galaxies[20]. However, only a few objects have been found in the transient stage of tidal stripping[21,22], and this assumed evolutionary path[19] has never been fully traced by observations. Here we show that 106 galaxies in the Virgo cluster have morphologies that are intermediate between normal, nucleated dwarf galaxies and single-component UCDs, revealing a continuum that fully maps this morphological transition, and fills the "size gap" between star clusters and galaxies. Their spatial distribution and redder color are also consistent with stripped satellite galaxies on their first few pericentric passages around massive galaxies[23]. The "ultra-diffuse" tidal features around several of these galaxies directly show how UCDs are forming through tidal stripping, and that this evolutionary path can include an early phase as a nucleated ultra-diffuse galaxy (UDG)[24,25]. These UCDs represent substantial visible fossil remnants of ancient dwarf galaxies in galaxy clusters, and more low-mass remnants probably remain to be found.**

We performed a detailed structural analysis (Methods) of compact stellar systems in the nearby Virgo cluster of galaxies based on the Next Generation Virgo Cluster Survey (NGVS). Two intermediate populations were identified between typical dE,Ns and UCDs. First, we have selected 51 nucleated dwarf galaxies (dE,Ns) that have "overweight" central nuclear star clusters ("strongly nucleated" dEs), with a nuclear-to-total luminosity fraction of $f_{nuc} > 8\%$, that are among highest ~10% in $f_{nuc}$ for ~600 Virgo dE,Ns with $g'$-band absolute magnitude $M_g$ between −16 and –9.5 mag. Second, we found that ~15% Virgo UCDs[4] cannot be described by a single King or Sérsic profile with reasonable parameters, indicating the existence of extended envelopes. Follow-up Gemini Multi-Object Spectrograph (GMOS) spectroscopy (Methods), together with previous observations[26,27], brings to a clean sample of 55 UCDs with envelopes (eUCDs) in Virgo, of which 50 were selected with NGVS data (including a few bright eUCDs identified in previous HST work[8,14,15]), and an additional five eUCDs that could only be resolved in the Hubble Space Telescope (HST) imaging from the ACS Virgo Cluster Survey (ACSVCS) (Methods). Our sample fully bridges the morphological gap between dwarf galaxies and UCDs, as shown in Fig. 1a, despite the fact that eUCDs are much more compact than strongly nucleated dEs, and were not initially discovered as galaxies.

Are these strongly nucleated dEs and eUCDs truly transition objects? The imaging depth (surface brightness $\mu_{g'} \approx 29$ mag arcsec$^{-2}$) of both NGVS and the Burrell Schmidt Deep Virgo Survey (BSDVS)[28] has enabled us to visually identify significant S-shaped tidal features and streams—unique indicators of the ongoing tidal disruption[29]—around a few strongly nucleated dEs and eUCDs (Fig. 1b, see Methods for more details). Only a few Virgo galaxies with similar features have been reported before[21,28,30]. We find that VCC 1672 near M89 and NGVS J123037.24+124609.2 (hereafter NGVS 2078) near M87 both show tidal features at surface brightnesses $\mu_{g'} > 27$–28 mag arcsec$^{-2}$ that span more than 3 arcmin (about 14 kpc). We also find the approximately 20 kpc-long tidal streams associated with eUCD NGVS-UCD156 (near NGC 4365) and the bright globular cluster NGVS J122846.19+134311.4 (hereafter NGVS J1229+1343)[4], and a tentative 70 kpc stream originating from eUCD NGVS-UCD509 near M87. Locations and zoom-in images of tidally disrupting galaxies in the Virgo core region are highlighted in the Fig. 1b. Their proximity to massive galaxies, and resemblance to the distorted morphologies shown in numerical simulations[23,29], strongly suggest ongoing tidal disruption rather than imaging artefacts. These new findings now bring the number of known disrupting dE,Ns in Virgo into the range predicted by simulations[31].

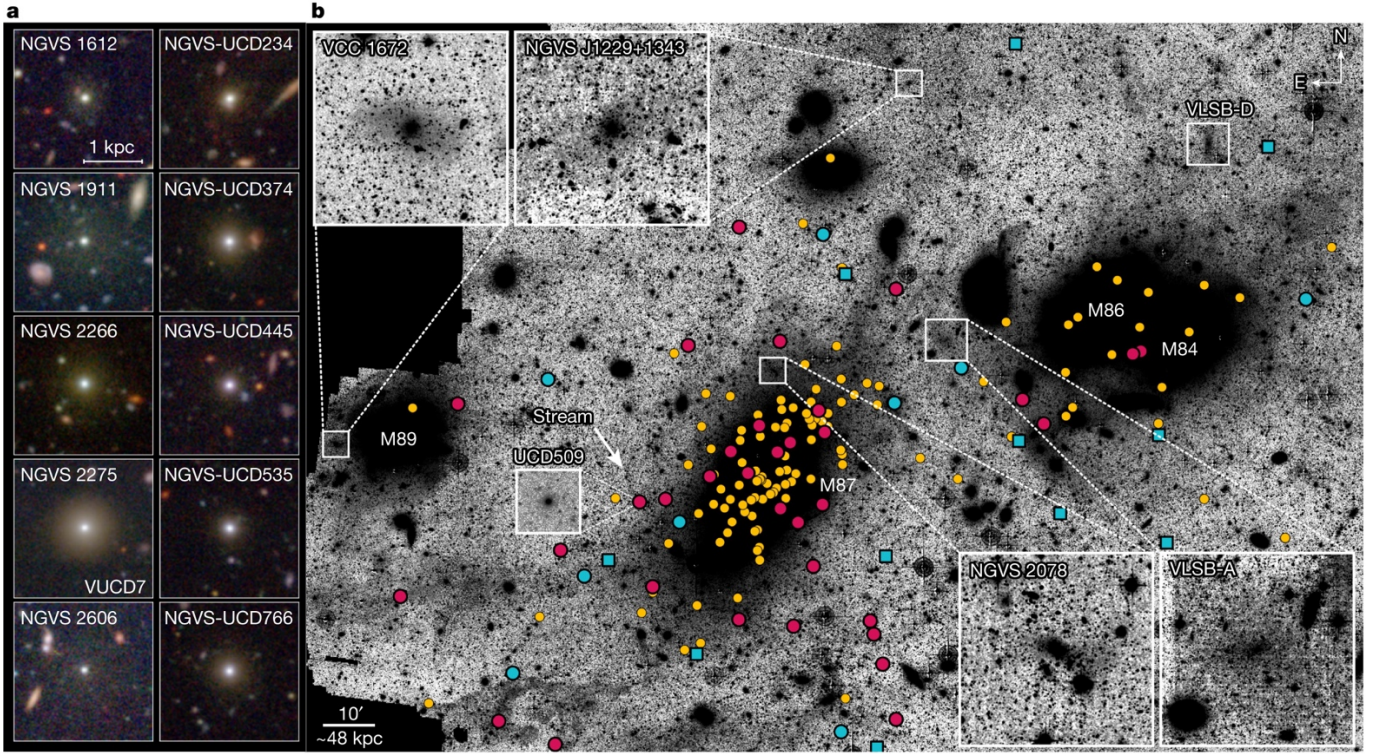

**Fig. 1. | Strongly nucleated dEs, eUCDs and UCDs in the Virgo Cluster core. (a)** $u^*g'i'$ color-composite images of representative strongly nucleated dEs (left column) and eUCDs (right column). Each cutout image is 0.5 arcmin × 0.5 arcmin. **(b)** BSDVS imaging of low-surface brightness features in the Virgo core region. Yellow and dark red circles represent UCDs and eUCDs, respectively. Blue circles and squares are strongly nucleated dEs with and without velocity measurements. $u^*g'i'$ co-added and Gaussian smoothed deep NGVS imaging of four disrupting dE,Ns (VCC 1672, NGVSJ 1229+1343, NGVS 2078, VLSB-A) and NGVS-UCD509 are shown in the insets. The arrow indicates the 70 kpc-long tidal stream that likely originates from NGVS-UCD509.

On the size-luminosity scaling relation (Fig. 2), most strongly nucleated dEs with $M_{g'} < -11$ are actually larger and thus more diffuse in surface brightnesses relative to the main locus of Virgo dE,Ns, suggesting tidal interaction and heating to puff up galaxies. Only the fainter ($M_{g'} \approx -10$) strongly- nucleated dEs have sizes consistent with their normal dE,N counterparts. Fig. 2 also shows the sizes of eUCD envelopes span approximately 30−200 pc, filling the long-observed size gap between star clusters and galaxies[32]. Strongly nucleated dEs and eUCD envelopes map out a contiguous sequence in decreasing size and luminosity, something which is expected by repeated episodes of tidal stripping at each pericentric passage of a massive host[19].

Those dE,Ns undergoing active tidal disruption (VCC 1672, NGVS 2078, NGVS J1229+1343, and VLSB-A) show more complicated structures, which includes a compact inner stellar envelope with effective radius $R_e$ of a few hundred parsecs surrounding the central nuclear star cluster, and an elongated and twisted outer stellar component or stream with $\langle\mu\rangle_{e,g'} > 27$ mag arcsec$^{-2}$. Such "tidal breaks" in surface brightness profiles, which has been predicted in simulations[33], requires a three-component model to describe (Methods), rather than the classical two-component "nucleus+main galaxy" model for undisturbed dE,Ns. The size-luminosity diagram shows that the main galaxy (inner+outer) components (pentagons in Fig. 2) occupy the same parameter space as Virgo UDGs[30], while the central nuclear star clusters (NSCs) (circles in Fig. 2) are indistinguishable from UCDs and massive globular clusters (GCs). However, if the outer component is entirely removed (or is below the surface brightness detection limit), the inner galaxy component alone (diamonds in Fig. 2) is structurally similar to strongly nucleated dEs or eUCDs. These components are connected in Fig. 2 and provide details about the potential evolution in the size-luminosity diagram, ending up as UCDs.

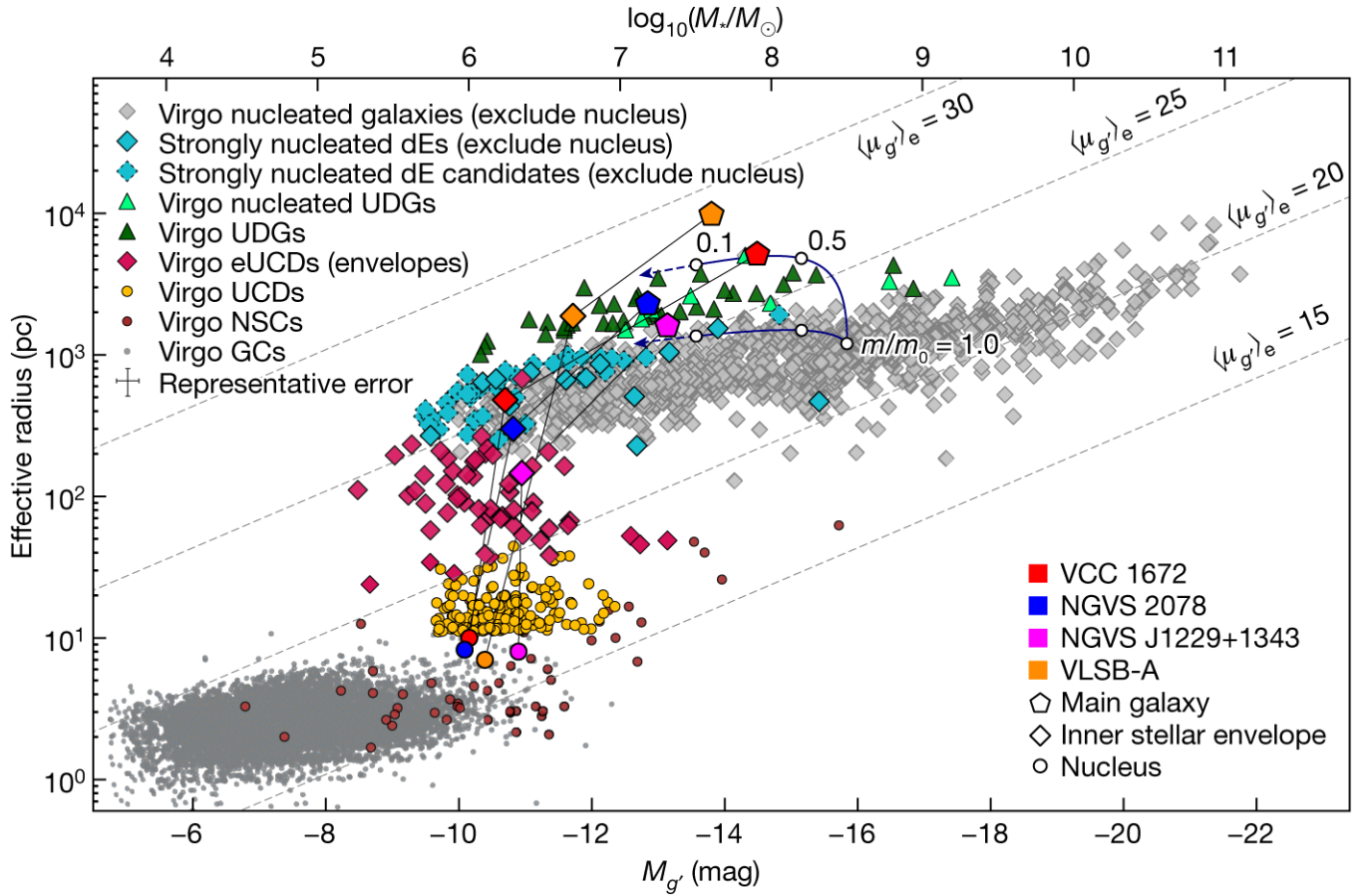

**Fig. 2. | The size-luminosity scaling relationship of Virgo early-type stellar systems.** The stellar envelopes of strongly nucleated dEs (blue) and eUCDs (dark red) form a distinct population when compared to other Virgo nucleated galaxies (gray diamonds) from NGVS[48], with eUCDs filling the size gap between normal galaxies and star clusters[4,49,50]. Some strongly nucleated dEs have sizes comparable to UDGs (dark/light green for non-nucleated/nucleated UDGs[30]). Tidally disrupting dE,Ns are decomposed and each labeled with three connected points representing each component. The evolutionary tidal stripping track[34] for galaxies in cored (upper) and cuspy (lower) dark matter halos are shown as dark blue curves with arrows denoting where the track is extrapolated. The remaining mass fraction, compared to the initial stellar mass ($m/m_0$), is labelled along the track. Dashed lines from lower-left to upper-right denote constant mean effective surface brightness. A representative $1\sigma$ error bar is shown in the legend.

Previous simulations[33,34] show that the change in stellar mass and half-light radius of a tidally stripped galaxy can be parameterized as a function of the total amount of mass lost within the half-light radius and the dark matter halo density profile. Fig. 2 plots these evolutionary tracks of total stellar mass and half-light radius for a typical satellite ($R_e \approx 1$ kpc, $M_* \approx 10^{8.5} M_\odot$) that is embedded in a cored (upper track) or cuspy (lower track) dark matter halo profile. If the progenitors were dEs, as hypothesized here, cored dark matter halos may more easily produce an evolutionary stage with large and diffuse envelopes, consistent with UDGs[25]. More extended disky progenitors would be needed to produce sufficient UDGs in the case of cuspy dark matter profiles[35]. The diversity in size and surface brightness of strongly nucleated dEs suggests that their dark matter halo profiles might differ. dE,Ns possessing very diffuse stellar envelopes that are comparable to ultra-diffuse galaxies (UDGs[24]) are more likely to be embedded in cored dark matter halos[25], while cuspy dark matter halos can account for those more compact and higher surface brightness envelopes. The existence of very diffuse disrupting dE,Ns provides more direct evidence that tidal stripping contributes to the production of both the UCD and UDG populations[21]. The evolutionary pathway to forming UCDs may sometimes include a transient stage as a nucleated UDG, which through stripping would also be dark matter deficient[36,37].

Tidal stripping drives the increase of the nuclear-to-total luminosity fraction ($f_{nuc}$), as NSCs are much more resilient to tidal stripping and will remain mostly unchanged until the entire galaxy is fully disrupted[38]. For a typical dE,N ($f_{nuc}$ = 1%) to evolve into one of our strongly nucleated dEs ($f_{nuc} \approx$ 8%−35%) via tidal stripping, it would lose over 90% of its stellar mass. To form a eUCD with even higher $f_{nuc}$ (about 30%−80%), it will have lost about 99% of its stellar mass. For the average eUCD ($\langle g'_{tot}\rangle_{eUCD}$ = 20.0 mag and $\langle f_{nuc}\rangle_{eUCD}$ = 56%), this corresponds to a typical dE,N progenitor of approximately $10^8 \, M_\odot$.

Compact stellar systems are predominantly more metal-rich and redder than typical galaxies at the same stellar mass[39]. Fig. 3 shows the color-magnitude diagram for stellar envelopes of eUCDs and strongly- nucleated dEs, compared to the inner color (within effective radius) of Virgo early-type galaxies[40]. The mean envelope ($g'$−$i'$) colors of eUCDs and strongly nucleated dEs are respectively ~ 0.07 and 0.03 mag redder than galaxies at the same magnitude, which is similar to the colors of dE,Ns with 17 < $g'$ < 19 mag ($M_* \sim 10^{7.5} \, M_\odot$), 2−4 magnitude brighter than the bulk of eUCDs. This provides independent support for the idea that these eUCDs started out as galaxies ~10−40× more massive, with more normal $f_{nuc}$. The smaller red offset for the strongly nucleated dEs is consistent with their lower $f_{nuc}$, both of which indicate that these galaxies have not undergone as much transformation as the eUCDs. A fully stripped dE,N will contribute its NSC to the large population of UCDs ($g'$ = 20−21 mag) or globular clusters ($g' \sim$ 22−23 mag or fainter). We found that the nuclear color of eUCDs and bright strongly nucleated dEs shows no difference relative to the color-magnitude relation of galaxy NSCs[41] (Extended Data Fig. 7).

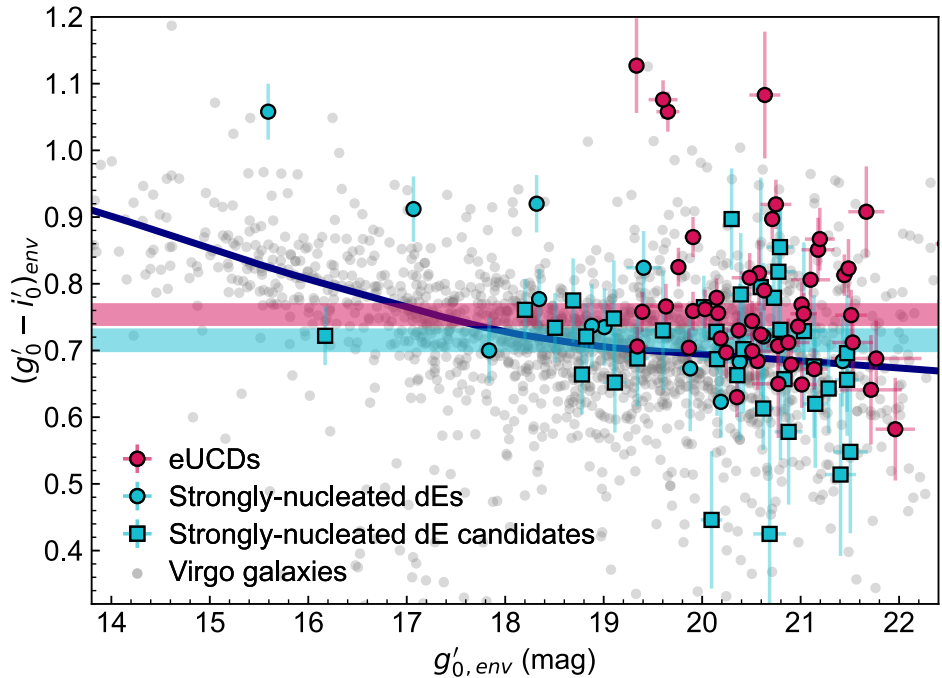

**Fig. 3. | Color-magnitude diagram for eUCDs, strongly nucleated dEs and Virgo early-type galaxies.** The stellar envelopes of eUCDs and strongly nucleated dEs are redder than their dE counterparts at the same luminosity. The mean envelope ($g'$−$i'$) color of eUCDs (red horizontal band) is ~0.07 mag redder, and

the mean envelope *(g′−i′)* color of strongly nucleated dEs (blue horizontal band) is ∼0.03 mag redder, than the dwarf galaxy color-magnitude relation (dark blue line) at the same magnitudes. The thicknesses of the red and blue bands represent the 1$\sigma$ uncertainties in the mean envelope color for eUCDs and strongly nucleated dEs, respectively. Where these bands intersect the color-magnitude relation shows what the mean progenitor luminosity for each population is expected to be, assuming evolution due to tidal stripping. A 1$\sigma$ error bar for the color and magnitude is shown for each object.

The structural properties and colors of strongly nucleated dEs and eUCDs suggest that they are a transitional phase between dE,Ns and UCDs, with tidal features providing compelling evidence that tidal stripping is the linking physical mechanism. In Fig. 4, we show what an evolutionary sequence due to tidal stripping would look like in observations. The sequence provides a panorama of dE,Ns at different stages of such transformation, from a normal dE,N to a UDG, then an eUCD (e.g. NGVS-UCD509), and finally to a "naked" UCD/GC (e.g. NGVS-UCD507).

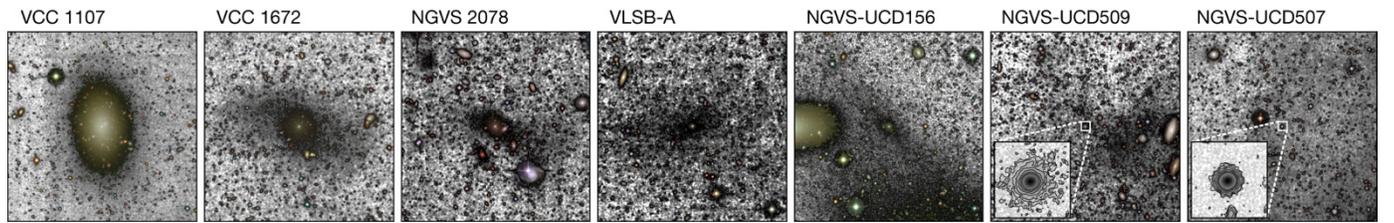

**Fig. 4. | Morphological Sequence of tidal stripping.** Each 3.75 arcmin × 3.75 arcmin panel shows a dE,N or a UCD, ordered by increasing nuclear luminosity fraction. The sequence starts with VCC 1107, which is an undisturbed dE,N at the starting point of the evolutionary track in Fig. 2, and the subsequent galaxies all have similar NSC magnitudes. A zoom-in with contours for eUCD NGVS-UCD509 and UCD NGVS-UCD507 is shown in the bottom-left corners to illustrate their different morphology. VCC1672, NGVS 2078, and VLSB-A all meet the structural definition of a UDG.

A history of recent interactions for strongly nucleated dEs and eUCDs is also evident from their strikingly different spatial distributions compared with "naked" UCDs around M87 (Fig. 1b). The distribution of strongly nucleated dEs and eUCDs is less concentrated and much more inhomogeneous than the bulk of confirmed single component UCDs, as well as GCs[42]. Strongly nucleated dEs and eUCDs both trace shell-like patterns and aligned linear substructures around M87 (also see Extended Data Fig. 6), mostly in the outer regions of the galaxy, similar to previous numerical simulations[18,31]. Considering the typical formation timescale of eUCDs in the dense cluster environment[18], this spatial distribution suggests that a group of galaxies has been accreted and fallen into the cluster center in the past 2−3 Gyr[43,44].
The nucleation fraction of dwarf galaxies with $M_*$ between approximately $10^8 - 10^{10}$ $M_\odot$ is higher than 70%[17], so UCDs represent the majority of disrupted normal nucleated dwarf galaxies in this mass range, pointing to a steeper galaxy luminosity function in the past[45]. These UCDs may also host a significant number of massive black holes in our local Universe[46]. Finally, disrupted dE,Ns contribute to the intra-cluster light[28] (ICL) and stellar halos of massive galaxies. We estimate the stellar mass contribution from 16 strongly nucleated dEs, 33 eUCDs and 89 confirmed UCDs in 1 degree around M87. Assuming all of these objects are stripped nuclei whose progenitors had $\langle f_{nuc} \rangle$ = 1%, they will contribute approximately 7 × $10^{10}$ $M_\odot$, which is ∼ 30% of the stellar mass within the M87 stellar halo outside of 20 kpc[45,47]. Given that evolved tidal streams are extremely faint (usually fainter than 30 mag arcsec$^{-2}$), we expect that future deep imaging from Rubin Observatory Legacy Survey of Space and Time and the Nancy Grace Roman Space Telescope will reveal more tidal streams around these transitional objects, and fully illuminate their evolutionary connection.

## Methods

### NGVS

The Next Generation Virgo Cluster Survey (NGVS)[51] is a deep, high-resolution multi-wavelength ($u^*$, $g'$, $r'$, $i'$, $z'$) imaging survey of the entire Virgo cluster conducted with MegaCam on Canada–France–Hawaii Telescope (CFHT/MegaCam[52]). The survey footprint covers 104 deg$^2$ within the virial radii of the Virgo A and Virgo B subclusters, centered on M87 and M49, respectively. In the $g'$-band, the point-source detection limit is 25.9 AB mag ($10\sigma$), and the extended-source detection limit is 29.0 AB mag arcsec$^{-2}$ ($2\sigma$ above the sky). Throughout the paper, we assume our objects locate at the Virgo distance of 16.5 Mpc, unless the host galaxies can be clearly determined and the distance of the hosts are independently known[53].

Taking advantage of the excellent image quality ($i'$-band seeing less than 0.6″, and about 0.8″–1.0″ for other bands), Liu et al.[4] presented a sample of 612 confirmed and candidate UCDs in the Virgo Cluster using a combination of half-light radius ($11 < r_h < 100$ pc), luminosity ($g' < 21.5$ mag), morphology, colors, and available radial velocity information. Apart from the UCDs that were already spectroscopically confirmed (UCDs with radial velocity $v_r < 3500$ km/s, about ~1/3 of the sample), an additional 88 UCD candidates in the catalog were selected on the basis of deep $u^*g'i'z'K_s$ data to have a high probability of being Virgo members. The background contaminant fraction is expected to be higher in the outskirts of the Virgo cluster, especially in the northernmost region ($\delta > 16$ deg), where only $u^*g'z'$ color selection is available. Visual inspection by Liu et al. identified 41 objects with potential envelopes, including 29 confirmed UCDs and 12 UCD candidates, although no structural fitting or photometry of these envelopes were performed. Strongly nucleated dEs that could potentially be UCD progenitors are selected in the NGVS galaxy catalogue[47], which lists a total of 3689 galaxies in the Virgo cluster. Note that none of the objects in the UCD samples, including eUCDs, are included in the dE,N sample derived from the full NGVS galaxy catalogue. VUCD7, which is clearly brighter than most UCDs, is included in the strongly nucleated dE sample (NGVS 2275) as it was identified as a dwarf galaxy in the NGVS, being quite extended in appearance. Among the 586 early-type dwarf galaxies that have a total magnitude $15 < g' < 21.5$ ($-16 < M_{g'} < -9.5$) and NSC $R_e < 0.5$ arcsec (40 pc, to remove spurious nuclei), 51 dE,Ns have $f_{nuc} > 8\%$. Galaxies that are identified as compact elliptical galaxies (cEs)[54] were not included in this study. Although nuclear star clusters (NSCs) are prevalent in nearby galaxies of any morphological type[17], and stripped nuclear star clusters may come from spiral galaxies, no strongly nucleated spiral galaxies were found in Virgo. We expect that processes such as tidal interaction and ram-pressure stripping may have transformed those late-type galaxies to early-type galaxies during the formation of UCDs, thus such strongly nucleated spiral galaxies may be rare in dense environments. We focus on the dwarf elliptical galaxies as they could better show the evolutionary sequence and compare their properties with our UCD sample, such as size, surface brightness, and colors.

While the NGVS data has both excellent image quality and deep surface brightness sensitivity, we are still limited by the data in both respects. The ground-based spatial resolution of CFHT limited the parent sample of 612 UCDs to a size cut of $r_h > 11$ pc. This cut removes nearly all foreground stars, but also removes UCDs that have smaller sizes (as well as many globular clusters). Although more typical nuclear star clusters are known to have half-light radius $< 10$ pc[17], this falls below the limit for reliable size measurements at Virgo distance, which causes the artificial size discontinuity between UCDs and normal GCs seen in Fig. 2. Fortunately, UCDs with envelopes can have single-component King model $r_h > 11$ pc, even if the decomposed nucleus is smaller. In that way, our data is able to probe to smaller nuclear size scales, even if the NSC itself is unresolved (but the envelope is resolved). The NGVS data are thus best suited to finding objects that larger, and have brighter and larger envelopes, i.e., those in the earlier phases of tidal stripping. However, it is difficult to decompose UCDs that have large ($r_h > 20$–30 pc) and bright NSCs with compact envelopes (for example, M60-UCD1) using only NGVS data. This work therefore presents a lower limit on the number of visible stripped nuclei in the Virgo cluster, with fainter and smaller stripped nuclei remaining to be found with space telescopes. As a hint toward what remains to be found, eight bright globular clusters with $g' < 21$ mag also show diffuse envelopes (ref.4, table A1), but are not included in the catalog of 612 Virgo UCDs due to their slightly smaller sizes or mismatched fitting results between $g'$ and $i'$ band.

## BSDVS
The Burrell Schmidt Deep Virgo Survey (BSDVS) was conducted with the Case Western Reserve University (CWRU)'s Burrell Schmidt telescope[21,28]. The final survey footprint consists of 15.3 deg$^2$ and 16.7 deg$^2$ in the Johnson $B$- and $V$-bands, respectively, with a pixel scale of 1.45″ pixel$^{-1}$. The survey covers the main mass concentrations (for example, the M87, M49, M86 & M84 regions) of Virgo, and is optimal for detecting very faint and extended features, including tidal features and intra-cluster lights (ICL). The $3\sigma$ limiting surface brightnesses are $\mu_{B,lim} = 29.5$ mag arcsec$^{-2}$ and $\mu_{V,lim} = 28.5$ mag arcsec$^{-2}$.

## ACSVCS
The ACS Virgo Cluster Survey (GO-9401)[55] is a program to image 100 early-type galaxies of the Virgo Cluster using the Advanced Camera for Surveys (ACS) on the HST in two widely separated bandpasses (F475W/$g'$ and F850LP/$z'$). There are 22 UCDs observed by ACSVCS with 0.1″ resolution (full width at half maximum), all within ~2 arcmin of their host galaxy centers[56].

**Tidal features**
Various tidal features are commonly found in galaxy cluster environments, but here we focus primarily on tidal features that reflect the disruption of nucleated dwarf galaxies and the formation of UCDs. We performed extensive visual inspections for tidal features around all strongly nucleated dEs and all Virgo UCDs on 4×4 binned and smoothed $g'+r'+i'$ NGVS imaging, and compared NGVS imaging with available BSDVS imaging, where there was BSDVS coverage. In addition to the previously reported VLSB-A[21] and VLSB-D[28], VCC 1672 and NGVS 2078 stand out with elongated S-shaped low surface brightness features in their outskirts. Visual inspection of all Virgo UCDs[4] revealed that NGVS-UCD156 is the only UCD associated with a prominent 20-kpc long tidal stream that can be seen in smoothed NGVS imaging. NGVS-UCD156 is in the NGC 4365 group, which is thought to be an independent group of galaxies at approximately 6 Mpc behind the Virgo Cluster. NGVS-UCD509 near M87 is likely associated with a 70 kpc-long, but extremely faint, bent tidal stream that is more visible on BSDVS imaging. The brightest southern part of this stream, located 48″ (approximately 38 kpc) southwest of NGVS-UCD509, is catalogued as NGVSUDG-17[30] and may be stripped material from the progenitor dE,N, which can be seen in Fig. 4. Both NGVS and BSDVS reveal the faint but noticeable over 20 kpc-long tidal stream that originates from a bright ($M_{g'} = -11.34$ mag) globular cluster NGVS J1229+1343 with a diffuse envelope, suggesting NGVS J1229+1343 is a remnant nuclear star cluster of a disrupting dwarf galaxy. Located near Markarian's Chain, NGVSJ 1229+1343 is at similar projected distances to NGC 4459 (16 arcmin ≈ 75 kpc), NGC 4477 (19 arcmin ≈ 90 kpc), and NGC 4473 (23 arcmin ≈ 110 kpc), so it is difficult to unambiguously identify an obvious host that may be causing the disruption of NGVS J1229+1343. Another Virgo nucleated UDG, NGVSUDG-01[30], is embedded in a very long, narrow filamentary tidal stream[57] that provides additional evidence for the UDG-UCD formation connection. The tidal features around one bright strongly nucleated dE, VCC 1148 (NGVS 1781), and its companion dwarf galaxy VCC 1153, have also been reported[28].

Most of the remaining strongly nucleated dEs (47/51) do not show signs of tidal features down to an imaging depth of $\mu_{g'} \sim 29$ mag arcsec$^{-2}$, but this does not preclude there having been a history of interactions. An orbital study[58] of one nucleated UDG (VCC 615) without tidal features suggests that it likely passed near the cluster center and thus experienced some degree of tidal stripping, but the resulting tidal features may be too faint to be detected with current NGVS and BSDVS imaging. The elongated GC distribution and dynamical mass measurement for some UDGs are also independent evidence for the tidal stripping[59]. Although we cannot rule out the possibility that some strongly nucleated dEs directly formed an over-massive NSC, especially at the faint end, where $f_{nuc}$ is generally higher than for dE,Ns at higher masses[60]. However, the continuum seen in $f_{nuc}$ and effective radius, and the very different spatial distribution for strongly nucleated dEs compared to more typical dE,Ns (Extended Data Fig. 6), suggests that these galaxies are part of an evolutionary sequence. Meanwhile, we noticed that many strongly nucleated dEs in the Virgo core regions lack GCs, which is also consistent with the tidal stripping scenario. If they were not formed via stripping, an explanation would be needed for the formation of such a massive central NSC, one containing more than 10%−40% of a galaxy's stellar mass.

**2D image decomposition and identification of eUCDs**
We used GALFIT[61,62] to fit multiple models to each UCD in 200 × 200 pixel (0.62 × 0.62 arcmin) sub-images: a single King[63] (K), a single Sérsic[64] (S), and two-component King+Sérsic (K+S), double Sérsic (S+S) or PSF+Sérsic (P+S) models. The effective point spread function (PSF) was generated by the Astropy package Photutils[65] using multiple nearby unsaturated and isolated stars. For objects that had both NGVS and HST imaging, we found that for the two-component models, the K+S were more consistent between the two data sets than S+S. The P+S models often had worse results, even in NGVS imaging, indicating that many NSCs are slightly resolved. For these reasons, we subsequently only discuss the K+S two-component fits.

The initial fitting parameters are based on the magnitude and half-light radius from previous measurements[4]. For the initial Sérsic index ($n$) and King concentration ($c = \log10(r_t/r_c)$, where $r_t$ is the tidal radius and $r_c$ is the core radius), we use typical values of $n$ = 2 and $c$ = 1.5, which are broadly representative of early-type galaxies and globular clusters, respectively. For the two-component fit, we adopt an initial guess of $f_{nuc}$ = 30%, and outer Sérsic $R_e$ = 100 pc and $n$ = 1.5. The input shape of all components is set to an axis ratio $q$ = 1, given the fact that most UCDs are fairly round. All model parameters are allowed to vary in the fitting, except a fixed shape parameter $\alpha$ = 2 for the King model, plus a constraint of Sérsic index between $n$ = 0.5−8. The x- and y-position are fixed to be the same for two-component fitting. The NGVS images are already globally sky-subtracted, and any residual local sky background and gradient was fitted and subtracted with GALFIT before fitting the UCD, which is performed after masking all objects in the image. When the UCDs are severely influenced by the non-uniform stellar light from nearby galaxies (about 20 UCDs in our 612 UCDs have this problem, but most of them are covered in ACSVCS), the neighboring galaxies were first modeled and subtracted, although the higher noise related to the bright sky background makes it more difficult to identify stellar envelopes in these objects. Generally, the single-component fit converged quickly, and residual images show that UCDs are well modeled and subtracted. The typical Sérsic index (n) for a single-component fit is $n$ = 1−4, with a median of $n$ = 2.58. The typical King concentration index for a single-component fit is $c$ = 1−2, with a median of $c$ = 1.54. A small but significant fraction (∼ 15%) of objects, however, cannot be reproduced well with a single profile. Under these circumstances, the Sérsic index usually reaches our upper limit of $n$ = 8, which implies that these objects may have extended envelopes. Objects that have a large Sérsic index in the single Sérsic model also show unphysical parameters in their single King model, including a runaway large tidal radius ($r_t$), large concentration index $c$ (e.g. $c$ ∼ 3), and very high central surface brightness. In these cases, two-component models provide a more physically meaningful description of their surface brightness profiles.

In this work, we address a potential structural continuum from objects with two clearly distinct components to those with a single component, and are always limited by the resolution of ground-based imaging. Although there is no fully objective or completely automated procedure to determine the existence of envelopes and select the best-fit model, we use a uniform methodology that is based on examining all the information from the data and models, plus our reasonable expectations for the structures of UCDs, case-by-case. Here we provide a summary of our classification of eUCDs. Our initial judgment is based on the single-component King and Sérsic model fits. We choose these models because, unlike e.g., power law profiles, they are physically motivated, always integrate to a finite luminosity, and are expected to fit a wide range of stellar dynamical systems. The single Sérsic index and single King profile concentration index served as first selection criteria for eUCDs. In practice, most typical eUCDs have Sérsic index $n$ > 6 and $c$ > 2 through our fitting. For these objects, we performed the two-component fit by randomly generating reasonable initial guess parameters multiple times for each model choice, as well as fine-tuning the parameters based on its 1D surface brightness profile, until we got the best-fit result. The goodnesses of the two-component fits are examined by comparing the reduced $\chi^2$ and residual maps with the single component fits. A good two-component fit is expected to subtract all the extended light excess, while a single component fit will still leave systematic residual features (e.g. outer rings). We further check the structural parameters of the best two-component fit result. These parameters should be reasonable, and consistent with the general properties of dE,Ns and UCD envelopes. We exclude objects if the component axis ratio $q = b/a$ < 0.3, or if the outer component has a higher central surface brightness than the inner component. For the rest of the UCDs, those that have a more realistic single Sérsic index and single King profile concentration index, we visually inspect their 1D surface brightness profile and

2D images, and search for any noticeable inflections in profile that are not caused by nearby sources. We then repeated the selection process above. The relative goodness between the one-component and two-component fit, and the choice of the best fit model, was also assessed using the Bayesian Information Criterion (BIC[66]). The integrated color information of the inner and outer components (shown in Fig. 3 and Extended Data Fig. 7) was derived by fixing the best-fit $g'$-band structural parameters and fitting only the magnitude in other bands. Examples of fitting results for eUCDs in Fig. 1 are shown in Extended Data Fig. 2.

All UCDs and UCD candidates are classified into four categories (0: No envelope, 1: Possible envelope, 2: Probable Envelope, 3: Certain Envelope) according to the prominence of stellar envelopes and fitting result, which are cross-checked by experienced team members. We found the trend in median difference of BIC scores between the one-component and two-component fits ($\Delta BIC = BIC_{one} - BIC_{two}$) is consistent with our classification, e.g., "certain" eUCDs tend to have the highest positive $\Delta BIC$, while $\Delta BIC$ of "possible" eUCDs are smaller and are more consistent with UCDs with no envelopes. We selected 84 objects to have at least "possible" envelopes (categories 1, 2, and 3), including 45 that were already confirmed to be Virgo members with radial velocity measurements from previous studies[26,27], and 39 UCD candidates that needed spectroscopic follow-up. We observed 38 of these 39 with Gemini North/GMOS in 2021 and 2022, and we confirmed 17 bona-fide Virgo UCDs (see Spectroscopy section). At this point, we have 50 "certain" and "probable" eUCD, and 12 out of 13 "possible" eUCDs, that are confirmed Virgo members, eliminating nearly all background galaxy contaminants. "Possible" eUCDs are typically UCDs that have a high Sérsic index or King concentration, but for which a stable two-component decomposition is difficult to obtain. In these cases, both the structural parameters and colors are expected to have significant uncertainties due to model degeneracy. For simplicity and consistency, we only plot the results of "certain" and "probable" eUCDs as eUCDs, and treat "possible" eUCDs as one-component UCDs throughout the paper. Not including the 13 "possible" eUCDs would not change any conclusions of our paper (spatial distribution, evolutionary sequence, etc). "Possible" eUCDs may represent more highly stripped nuclei that are evolving to "naked" UCDs. Finally, we analyzed 22 UCDs with HST imaging in the ACSVCS (see **Consistency Checks** below). Of these, five objects that were classified as "No envelope" using NGVS imaging were subsequently found to have envelopes when analyzed with the benefit of HST resolution. These five additional eUCDs already had measured velocities and were confirmed Virgo members, bringing our total sample of eUCDs to 55.

**Uncertainties on derived photometric parameters**

The main uncertainties in our eUCD decomposition include model mismatch, sky background subtraction, and PSF variation. Previous work has shown that GALFIT usually underestimates errors[67]. To get more realistic measurement errors, for each eUCD, we generate 100 realizations of mock eUCD images based on derived best-fit parameters, accounting for the Poisson noise associated with the source, background noise, and read noise that are calculated based on the properties in specific images. We also take into account the possibility of a systematic sky subtraction error by introducing a sky offset sampled from a Gaussian with a sigma of 3%, a value that likely overestimates the sky subtraction bias under our detailed analysis, but is used to be conservative. We then use GALFIT to repeat the decomposition for these mock UCDs. The $1\sigma$ uncertainties for size, magnitude, and color are calculated using the distribution of the fitted parameters for these mock observations. For the magnitude and color uncertainties we add in quadrature the derived systematic uncertainties for NGVS photometry[68]. A typical error bar for size and luminosity is shown in the Figure 2 legend. The typical magnitude error is around 0.15 mag, and the envelope size measurement error is around 15%. Color and magnitude uncertainties are shown for all strongly nucleated dwarfs and eUCDs in Figure 3. Although the uncertainties are unique to each object, and may be influenced by nearby neighbors or varying backgrounds, the typical uncertainties in both size and color are smaller than the scatter seen in the sample. We notice the color uncertainties of bright and compact UCD envelopes are less affected by sky subtraction error than those much more diffuse strongly nucleated dEs, for which the sky subtraction error is considered to be the dominat uncertainty.

**Structural decomposition of tidally disrupting dE,Ns**

We ran GALFIT on these galaxies to derive the structural parameters of the objects in $g'$-band. We performed accurate local sky subtraction using the GNU astronomy utility NoiseChisel[69], which detects extremely faint structures, allowing us to create masks of low surface brightness features in the image, including nearby galaxies and optical ghosts around bright stars, as well as artefacts like diffraction spikes. We then estimated the 2D local background map from unmasked regions using Photutils Background2D.

All nearby objects, including background sources overlapping with the target galaxies, are detected and masked using segmentation maps generated by Source Extractor[70]. Masks were created in two stages. The primary masks were created based on the original images. We fed the primary images into GALFIT to model and subtract the target galaxy. In the second stage, the residual map of the initial fit was used to create a second mask. The final mask for a given image is the union of these two masks. This process is done independently in $u^*$-, $g'$-, and $i'$-band images, and the final image mask is the union of the masks in all three filters. Any bright objects other than the smooth component and central nucleus of the galaxy are masked in the final image.

We took an iterative approach to our structural analysis. Instead of fitting multiple components simultaneously, we first fit the outer, extended component on co-added images. This component is most sensitive to accuracy of the background subtraction and the Sérsic index, so we initially fix the Sérsic index to $n = 1$ and fit for $R_e$ and the total magnitude of this component. All foreground and background objects, including nuclear components, were masked in this fit. We then performed an unconstrained second fit on the residual images, focusing on the profile of inner diffuse envelopes and nuclear components. Lastly, we put the parameters from extended components and nuclear components together and re-ran the fit simultaneously on images in multiple bands (allowing all parameters to be unconstrained) until we arrived at the best description of their overall profiles. The best-fit result for disrupting dE VCC 1672 can be found in Extended Data Fig. 2.

Using three components to decompose the structure of tidally disrupting dEs, rather than two com- ponents, is motivated, not only by the data, but also by theoretical and numerical studies that have suggested a more complex structure during tidal stripping[23,71,72]. Part of the stripped, unbounded stars will relax to virial equilibrium after the most recent pericentric passage, which causes a "tidal break"[33] that corresponds to the time elapsed since pericenter. Actually, simulations of UCD formation via tidal stripping also present a similar "tidal break" in the surface brightness profile during the early stages[19]. Moreover, previous observations[57, 73] of disrupting dE,Ns have already shown similar NSC + envelope + tidal stream structures. More complex structures are expected to be generated by multiple pericentric passages, which may help us to constrain the orbital history of these disrupting systems, e.g. VLSB-A shows a potential NSC + envelope + elongated bar-like structure + ultra-diffuse outer component. The stripped stars would eventually disperse with time, and the outermost component gradually becomes invisible. The net effect is consistent with the evolutionary sequence we observed in Fig. 4.

**Spectroscopy**

We obtained radial velocities for 9 dE,N candidates and 38 eUCD candidates using the Gemini Multi-Object Spectrograph (GMOS[74]) on the 8-meter Gemini North telescope during the 2021A and 2022A semesters. We used the 0.75 arcsec long-slit and B600 grating for 2021A, with a central wavelength at

5250 Å and a spectral resolution of $R \sim 1125$. For 2022A, we used the 1.00 arcsec long-slit, with a central wavelength at 6250 Å. For $g' = 21.5$ mag objects, the S/N per resolution element is $\sim 5$ with 45 min of exposure time.

We reduced the GMOS spectra using the PypeIt[75] package. After extracting the 1-D spectra, we derived radial velocities with penalized pixel fitting software pPXF[76,77]. We used high signal-to-noise (S/N $\sim$ 80) MMT/Hectospec stacked spectra of blue and red Virgo GCs as velocity templates[27]. We used the wavelength range $\lambda = 5500 - 6800$ Å, containing the strong $H_\alpha$ (6563 Å) absorption feature, which is among the most prominent spectral features in the relatively low S/N spectra we obtained. We identified 17 eUCDs as Virgo members with radial velocities $V_r < 3500$ km s$^{-1}$, and velocity uncertainties $\Delta v \sim 20 - 90$ km/s, and a median $\Delta v \sim 35$ km/s. The remaining 21 eUCD candidates are background galaxies, with redshifts ranging from $z = 0.028$ to 0.23. The high contamination fraction shows the key role of the spectroscopic survey in making an unbiased interpretation of properties of eUCDs. The bona-fide eUCDs and background galaxies confirmed in the spectroscopic survey show obvious differences in spatial distribution and shape. Background galaxies are mostly found in the low-density outskirt regions of Virgo, and all eUCD candidates with elongated ($b/a \lesssim 0.6$) shapes were found to be background galaxies. All confirmed eUCDs are round with ($b/a \gtrsim 0.7$). All nine observed dE,Ns are Virgo members, consistent with the fact that their surface brightnesses and sizes are similar to other Virgo galaxies. We expect that the contamination fraction for the rest of the strongly nucleated dEs without spectroscopy is also low. Example spectra for a bona-fide eUCD and a background galaxy are shown in Extended Data Fig. 3.

**Consistency checks**

We fitted 22 UCDs that are also covered in the ACS Virgo Cluster Survey[55] and compared the results with NGVS fitting. Empirical PSFs in the F475W and F850LP filters were derived using DAOPHOT II[78] using archival observations of fields in the outskirts of the Galactic globular cluster NGC 104 (47 Tucanae)[79]. For galaxies that have good sky subtraction in both NGVS and ACSVCS, we find their best-fit model and parameters are generally consistent, including effective radius, Sérsic index, and King concentration. For example, NGVS-UCD190, NGVS-UCD330, and NGVS-UCD395 are three eUCDs identified in NGVS imaging, which can be more easily decomposed into two components in HST imaging. The $f_{nuc}$ values derived from NGVS imaging are generally about 5% lower than those from HST imaging. The envelope component when fitting NGVS imaging tends to have a higher Sérsic index, and the effective radius is 10 − 20% larger than when fitting HST images. These behaviors can be explained by the shallower depth of HST imaging, which misses a small fraction of light in the outer envelope with surface brightness between $26 - 28$ mag arcsec$^{-2}$. This results in a smaller effective radius, lower Sérsic index, and a slightly higher $f_{nuc}$ estimation when using HST imaging. A comparison between HST imaging and NGVS image fitting for an eUCD is shown in Extended Data Fig. 4.

In addition to the eUCDs mentioned above that were in the ACSVCS, but could be identified in the NGVS as eUCDs, we were also able to use HST resolution to identify additional eUCDs with compact envelopes that were unresolved in the NGVS. Five UCDs in the ACSVCS have a single-component Sérsic index $n > 4$, in which three have $n > 7$ (NGVS-UCD167, NGVS-UCD386 and NGVS-UCD822). The two-component fits to these objects suggest that NGVS-UCD167 has a $R_e \sim 30$ pc envelope, and NGVS-UCD822 has a $R_e \sim 24$ pc envelope. While NGVS-UCD386 shows a power-law profile, and we are unable to decompose this object. Besides, NGVS-UCD192 has a prominent $R_e \sim 37$ pc envelope, and NGVS-UCD548 also show a two-component profile, with envelope size $R_e \sim 35$ pc. Another eUCD from HST imaging only is M60-UCD1 (NGVS-UCD753), which has a compact envelope ($R_e \sim 50$ pc), consistent with previous measurement[13]. Our NGVS work finds eUCDs with envelopes having $R_e > 40$ pc, which is likely a selection limit based on spatial resolution. These eUCDs with envelopes having $R_e < 40$ pc can only be resolved with HST imaging. A comparison between HST imaging and NGVS image fitting for an eUCD discovered only in HST imaging is shown in Extended Data Fig. 5. The eUCD-to-UCD fraction in NGVS imaging is over 12%, we suggest that this is a lower limit, and that the actual fraction should be higher. However, the current ACSVCS sample may be biased, in that the ACSVCS was a targeted galaxy survey, and so preferentially imaged areas where UCDs are more likely to form through interaction with a larger host. Future wide-area surveys of the Virgo cluster with HST, Euclid, or the Roman Space Telescope, will be needed to build an unbiased sample of UCDs/GCs with smaller envelopes.

## Data availability

NGVS data can be accessed via the Canadian Astronomy Data Centre (CADC). ACSVCS data is available via the Mikulski Archive for Space Telescopes (MAST) with the program ID HST GO-9401. Data products of Burrell Schmidt Deep Virgo Survey can be downloaded from http://astroweb.case.edu/VirgoSurvey/. Gemini/GMOS spectroscopic data of eUCD candidates taken during 2021-2022 can be downloaded through the Gemini Observatory Archive with program IDs GN-2021A-Q-208, GN-2022A-Q-206, and GN-2022A-Q-307.

## Code availability

We have made use of standard data analysis tools (GALFIT, Astropy, Photutils, Pypeit, NoiseChisel). All the software and code used in the research are accessible to the public.

## Acknowledgements

K.W. acknowledges support from the National Natural Science Foundation of China, grant No. 12273001, 12073002. C.L. acknowledges support from the National Natural Science Foundation of China, grant No. 12173025, 11833005, China Manned Space Project, grant No. CMS-CSST-2021-A04, and the MOE Key Lab for Particle Physics, Astrophysics and Cosmology. S.L. acknowledges the support from the Sejong Science Fellowship Program by the National Research Foundation of Korea (NRF) grant funded by the Korea government (MSIT) (No. NRF-2021R1C1C2006790). T.H.P. acknowledges support through FONDECYT Regular project 1201016 and CONICYT project Basal FB210003. C.S. acknowledges support from ANID/CONICYT through FONDECYT Postdoctoral Fellowship Project No. 3200959. E.T. is thankful for the support given by the NSF-AST- 2206498 grant. This work is based on observations obtained with MegaPrime/MegaCam, a joint project of CFHT and CEA/DAPNIA, at the Canada-France-Hawaii Telescope (CFHT), which is operated by the National Research Council (NRC) of Canada, the Institut National des Sciences de l'Univers of the Centre National de la Recherche Scientifique (CNRS) of France, and the University of Hawaii. This work is based on observations made with the Hubble Space Telescope. Support for program HST GO-9401 was provided through a grant from the Space Telescope Science Institute, which is operated by the Association of Universities for Research in Astronomy, Inc., under NASA contract NAS5-26555. Spectroscopic data were obtained at the international Gemini Observatory, a program of NSF's NOIRLab (Program ID: GN-2021A-Q-208; GN-2022A-Q-206; GN-2022A-Q-307). The international Gemini Observatory at NOIRLab is managed by the Association of Universities for Research in Astronomy (AURA) under a cooperative agreement with the National Science Foundation on behalf of the Gemini partnership: the National Science Foundation (United States), the National Research Council (Canada), Agencia Nacional de Investigación y Desarrollo (Chile), Ministerio de Ciencia, Tecnología e Innovación (Argentina), Ministério da Ciência, Tecnologia, Inovações e Comunicações (Brazil), and Korea Astronomy and Space Science Institute (Republic of Korea). Observations reported here were obtained in part at the MMT Observatory, a facility operated jointly by the Smithsonian Institution and the University of Arizona. MMT telescope time was granted by NSF's NOIRLab (Prop. ID: 2010A-0445, PI: E. Peng), through the Telescope System Instrumentation Program (TSIP). TSIP was funded by NSF. This work was enabled by observations made from telescopes (CFHT and Gemini North) located within the Maunakea Science Reserve and adjacent to the summit of Maunakea. We are grateful for the privilege of observing the Universe from a place that is unique in both its astronomical quality and its cultural significance.

## Author contributions statement

K.W. led the data analysis of the images and spectra of UCDs and nucleated dwarf galaxies, developed the main interpretation of the results, and wrote the manuscript. E.P. led the project design and management, the main interpretation of the results, and contributed to the manuscript. C.L. compiled the Virgo UCD samples and did exploratory work on the morphological sequence of dE,Ns and UCDs. J.C.M. led the Burrell Schmidt Deep Virgo Survey, discovered VLSB-A and VLSB-D, and contributed to the data analysis and the discussion of results. L.F. and P.C. led the NGVS and ACSVCS program design and management. L.F, P.C., and L.M. compiled the NGVS galaxy catalog. S.G. led the data production pipeline of NGVS data. A.L. and T.P. led the infrared observations of the NGVS. M.A.T and J.R. contributed to the Gemini spectroscopic observations. P.G. contributed to the observation proposals and the interpretation of the results. Y.K. provided the stacked MMT globular cluster spectra. All co-authors contributed to the discussion of the presented results and the preparation of the manuscript.

## Competing interest

The authors declare no competing interests.

## Correspondence

Correspondence and requests for materials should be addressed to Kaixiang Wang (kaixiang.wang@pku.edu.cn), Eric Peng (eric.peng@noirlab.edu), and Chengze Liu (czliu@sjtu.edu.cn).

## Additional information

**Extended data**

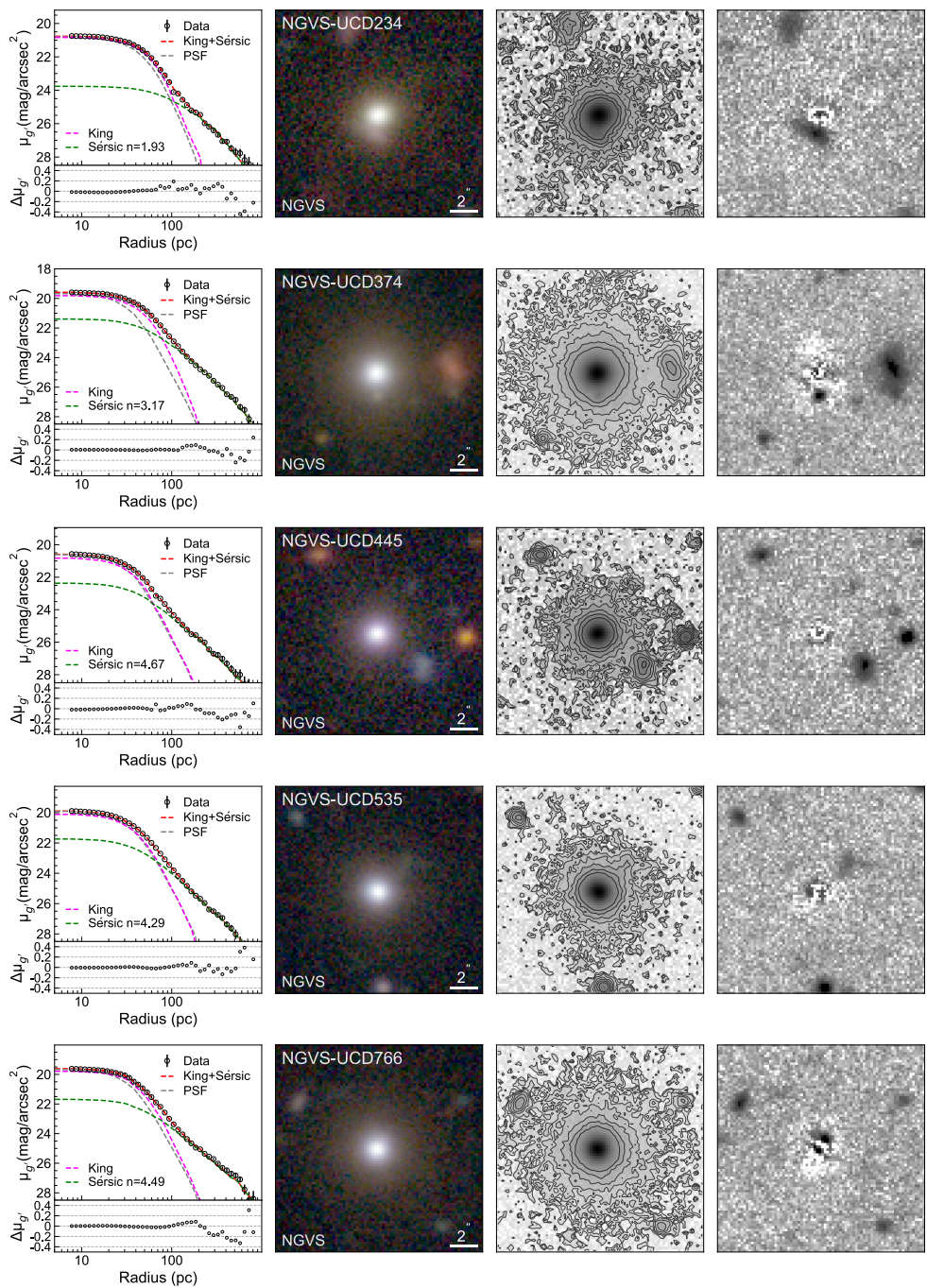

**Extended Data Fig. 1. 2D decomposition of eUCDs.** First column: Radial surface brightness profile of the UCD and best fit result of the two-component fitting. All eUCDs can be well-fit with an inner King model and an outer Sérsic model. The total, King, and Sérsic model profiles are shown with red, magenta and green dashed lines, respectively. The grey dashed line is the PSF profile. Error bars represent 1$\sigma$ uncertainties. The fitting residuals are shown below the source profile. Second column: NGVS $u^*g'i'$ color composite image. Third column: The $g'$-band image of UCD, with contours showing the surface brightness at $\mu_{g'}$ = 24 to 27 mag arcsec$^{-2}$. Last column: (data−model) residual map.

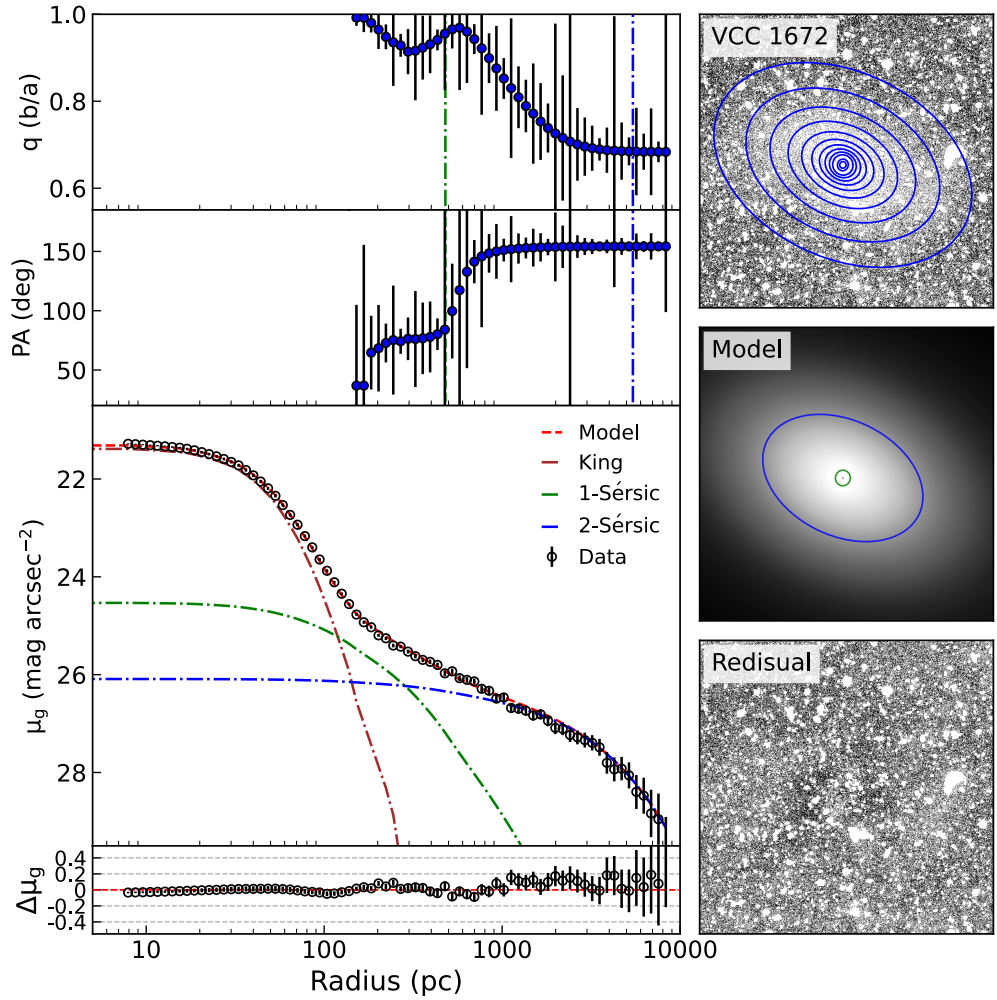

**Extended Data Fig. 2. The best-fit three-component model of VCC 1672.** The right panels display, from top to bottom, images of the data overlapped with isophotes, the best-fit model, and the residuals. North is up and east to the left. The left panels display the isophotal analysis of the 2D images and model fitting. From top to bottom, the panels show radial profiles of the axis ratio ($q$), the position angle (PA), the $g'$-band surface brightness, and the fitting residuals. For an illustration of the breaks in position angle and axis ratio, the effective radius of envelope components and outer components are plotted as vertical dot-dashed lines. Ellipses represent the effective radius of each individual component shown in the model image, which uses the same color legend as in the surface brightness plot. We note that the shifts in both $q$ and PA roughly correspond to the radius at which the outer component begins to dominate over the central envelope, at $R_e \sim 500$ pc. Error bars represent $1\sigma$ uncertainties.

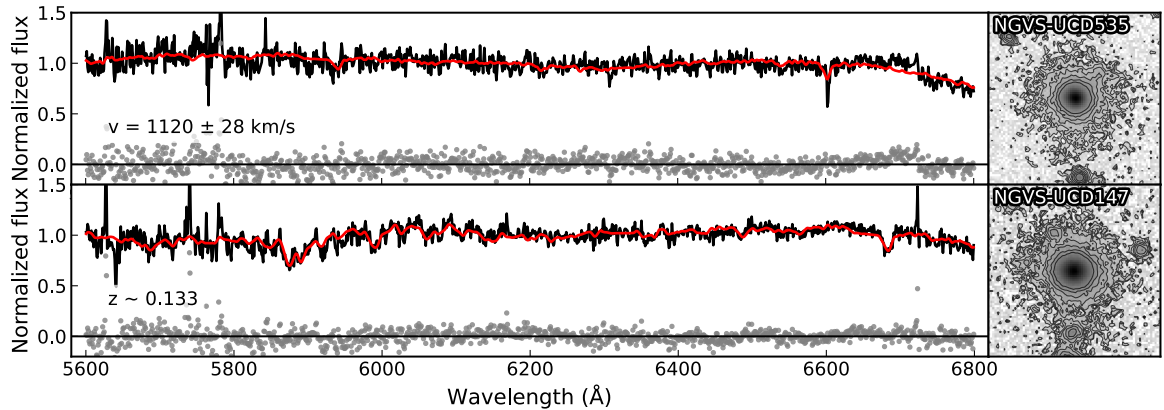

**Extended Data Fig. 3. Examples of Gemini/GMOS spectra of a bona fide eUCD and a background galaxy.** The Gemini/GMOS optical spectrum (black), the best-fitting template (red) and fitting residuals (gray) are plotted in the left panels. Right panels show the $g'$-band NGVS imaging and contours for each object, both of which show extended envelopes. The strong $H_\alpha$ absorption is the main diagnostic for bona-fide eUCDs. In contrast, the measured redshift of NGVS-UCD147 is $z \sim 0.133$, a background galaxy composed of bulge+disk.

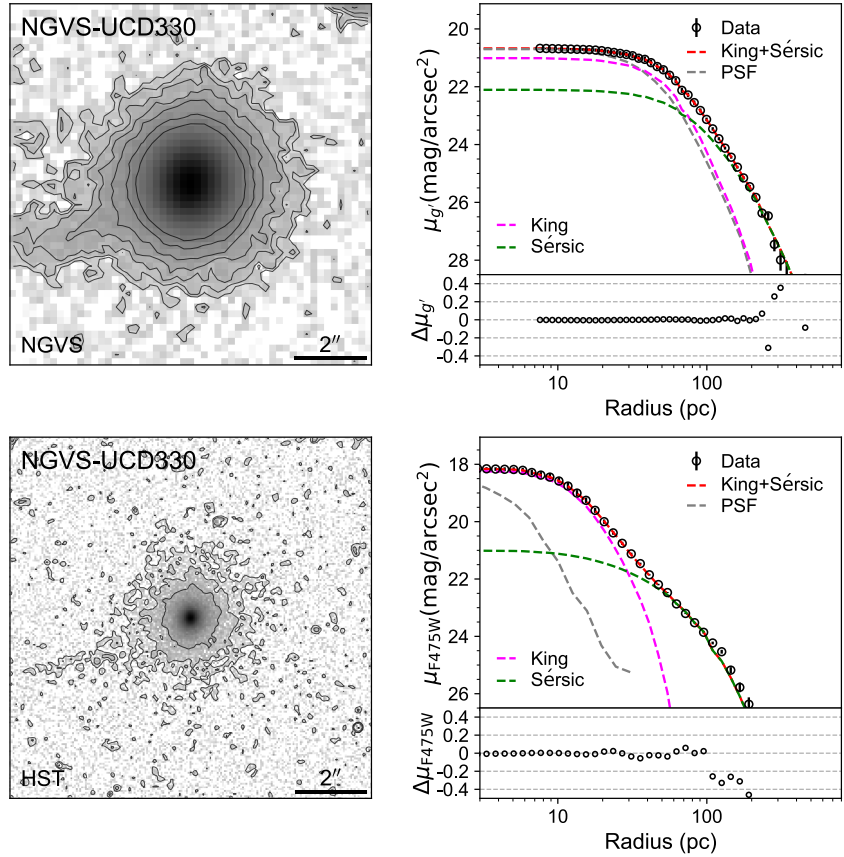

**Extended Data Fig. 4. A comparison between NGVS and HST imaging for NGVS-UCD330.** The upper and lower row are NGVS and HST data, respectively. Contours in the left column represent the surface brightness at $\mu_{g'} = 24-27$ mag arcsec$^{-2}$ in NGVS and $\mu_{F475W} = 23-25$ mag arcsec$^{-2}$ for HST. In the right column is the one-dimensional surface brightness profile of the UCD. NGVS-UCD330 shows marginal two components in NGVS imaging, and can be much clearly decomposed into two components in HST imaging. The total, King and Sérsic model profile are shown with red, magenta and green dashed line. The grey dashed line is the PSF profile. Error bars represent $1\sigma$ uncertainties.

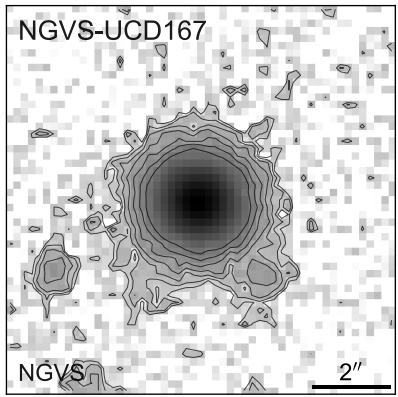
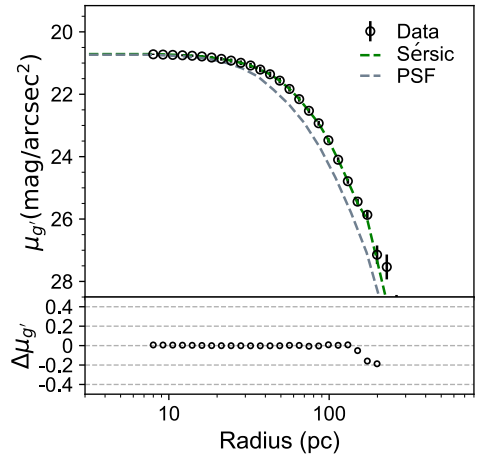
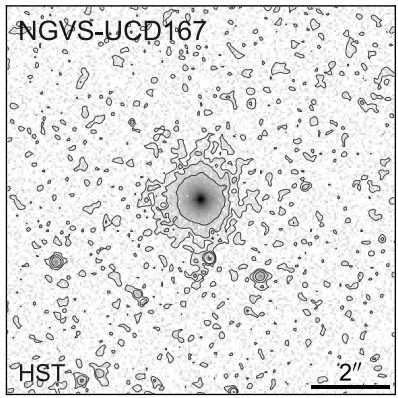
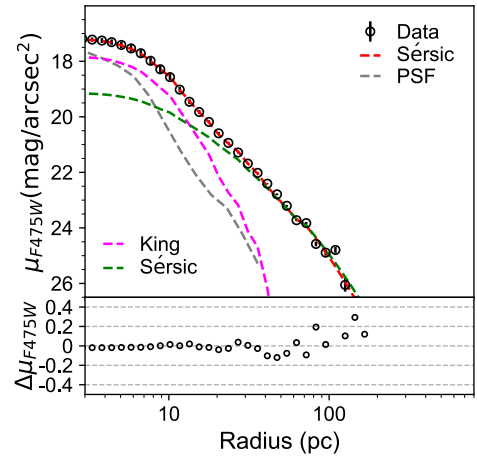

**Extended Data Fig. 5. A comparison between NGVS and HST imaging for NGVS-UCD167.** NGVS-UCD167 shows only one component in NGVS imaging, but can be resolved to two components in HST imaging.

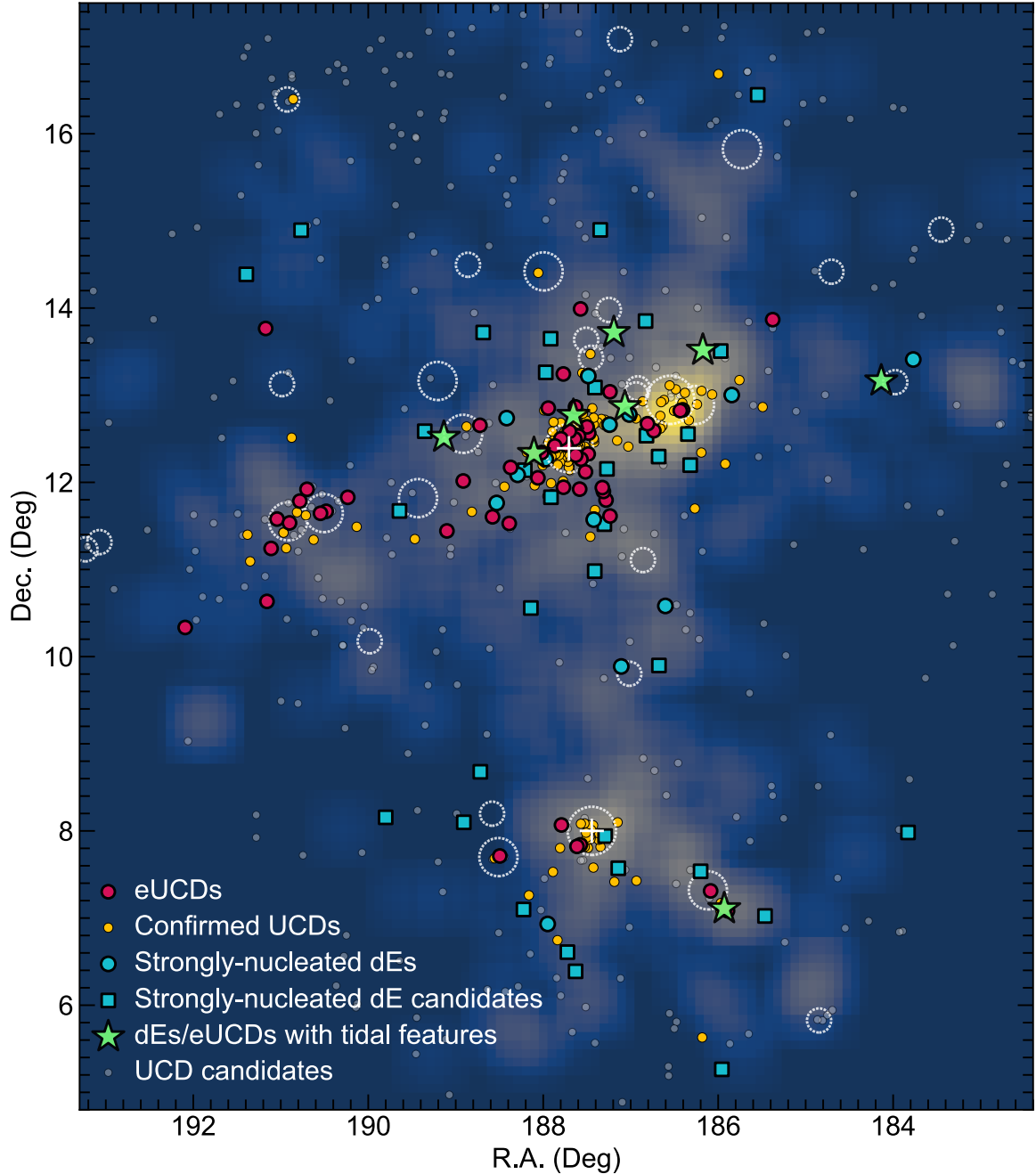

**Extended Data Fig. 6. Spatial distribution of compact stellar systems in the Virgo cluster.** Yellow and dark red circles represent UCDs and eUCDs, respectively. Blue circles and squares are strongly nucleated dEs with and without velocity measurements. Gray circles represent all UCD candidates without velocity measurements[4]. Disrupting dEs[21,28,30] and eUCDs that show tidal features are labelled with green stars. The Gaussian-smoothed distribution of Virgo dE,Ns identified in NGVS is shown in the background. While the density peak of all dE,Ns is located around M84 & M86 (northwest of M87), most eUCDs and strongly nucleated dEs are concentrated around M87. Massive Virgo galaxies ($M_{g'} < -20$) that could be primary contributors of tidal disruption are shown in white dashed circles, with circle sizes corresponding to their total luminosity. Most strongly nucleated dEs and eUCDs have clear associations with larger galaxies, or form coherent sub-structures.

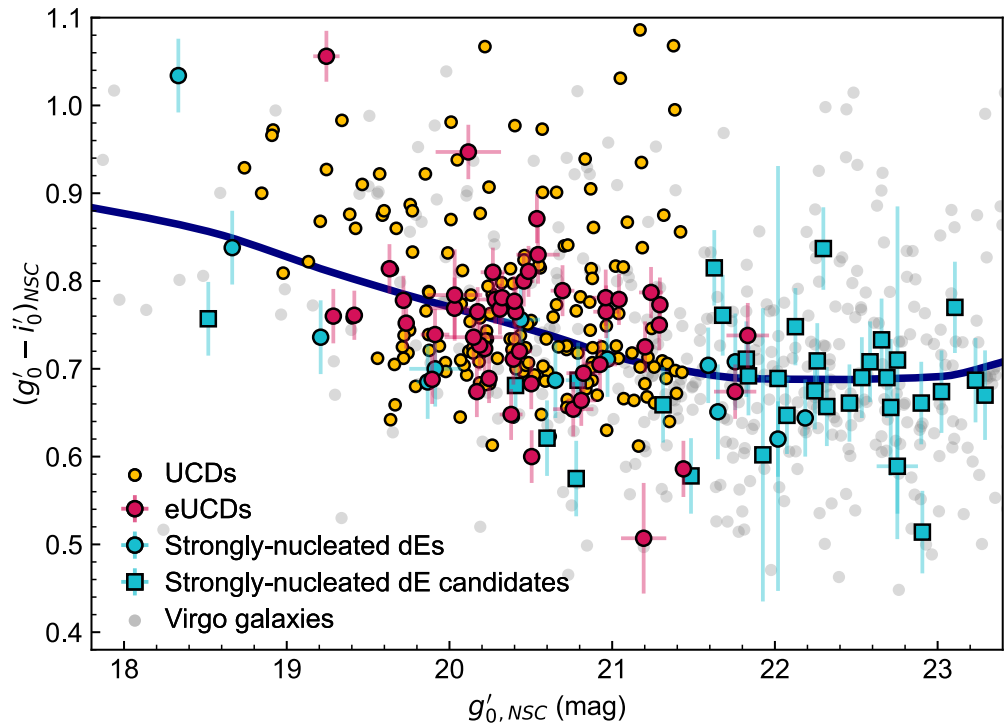

**Extended Data Fig. 7. Nuclear star cluster color-magnitude diagram for eUCDs, strongly nucleated dEs, and Virgo early-type galaxies.** Unlike the colors of the stellar envelopes, which are offset from the main galaxy color-magnitude relation, the nuclear star clusters in UCDs have colors that are very similar to UCDs without envelopes, and the NSCs of normal dE,Ns. $1\sigma$ error bar for the color and magnitude is shown for each object.